\documentclass[pra,preprint,superscriptaddress,showpacs,a4paper]{revtex4}
\usepackage{amssymb,amsmath,amsfonts}
\usepackage{graphicx}

\begin{document}

\title{Ionization of the hydrogen atom by intense ultrashort laser pulses}
\date{\today}

\author{S. Borb\'ely}
\affiliation{Babe\c{s}-Bolyai University, Faculty of Physics,400084 Cluj, 
Kog\u{a}lniceanu str.  nr. 1, Romania, EU}
\author{K.\ T\H{o}k\'{e}si}
\affiliation{Institute of Nuclear Research of the Hungarian Academy of 
Sciences, (ATOMKI), H--4001 Debrecen, P.O.Box 51, Hungary, EU}
\author{L. Nagy}
\affiliation{Babe\c{s}-Bolyai University, Faculty of Physics,400084 Cluj, 
Kog\u{a}lniceanu str.  nr. 1, Romania, EU}

\begin{abstract}
The ionization of atomic hydrogen in intense laser fields is studied
theoretically.
The calculations were performed applying both quantummechanical and
classical
approaches. Treating the problem quantummechanically, the time dependent
Schr\"odinger equation (TDSE) of our system was first transformed into a pseudo-momentum
space and solved in this space iteratively. While neglecting the Coulomb potential
during the solution of the TDSE we got the results
in the  Volkov approximation, in the first order solution we taken into account the
Coulomb potential
as perturbation. The classical calculations were performed within the
framework of the classical trajectory Monte-Carlo (CTMC) method.

The double differential ionization probabilities are calculated for
different laser pulses and a reasonable agreement was found between the theories. Major
differences can be observed in the angular distribution of electrons at low electron
energies between classical and the quantummechanical approaches. At high electron energies
the differences disappear, which indicates that the generation of low energy
electrons is of quantum type, and it is strongly influenced by the Coulomb potential, while
the production of high energy electrons is of classical type and it is less
influenced by the Coulomb interaction. Our results are also compared with the
Coulomb-Volkov (CV) model calculations. 
\end{abstract}

\maketitle

\section{Introduction}

In the last years advanced laser facilities have achieved intensities of 
the order of $10^{15} W/cm^2$ and pulse lengths of the order of 10 fs., which corresponds 
to few cycles of an electrical field of 800 nm wavelength \cite{nisoli, brabec00}. The process of interaction of such short and strong pulses with matter is a topic which has attracted much interest recently. 

In the multiphoton regime, many experimental \cite{agostini, paulus02, hansch, nandor, wiehle} and theoretical studies \cite{mainfray, parker, proto, radha} have 
been performed. In the tunneling regime, on the other hand, recent experiments \cite{rudenko, maharjan} with 
linearly polarized lasers have shown structures in the momentum distribution of the 
photoionized electrons in rare gases which have not been fully understood so far. The role of interference in few-cycle pulses is actually been both theoretically and 
experimentally investigated \cite{paulus01, lindner, arbo1}.

Moreover, the combined laser and charged particle induced electronic
processes either in ion-atom or ion-surface collisions are of considerable interest
both for basic and applied sciences \cite{kirchner}. From the fundamental point of view 
they might broaden our general understanding of the dynamics of atomic processes
for laser-matter interactions and field-free collisions.
As for the applications these studies can help us to
find the way for the control of ultrashort quantum processes
which are important in a number of applications, like in laser-driven fusion, in plasma heating, or in the
development of fast optical electronic devices.
The dynamics of atomic processes for the above mentioned interactions are not fully understood due to the lack of the exact and efficient theoretical models. For a detailed understanding of processes involved in the laser-matter interaction one needs to solve the time dependent Schr\"odinger equation (TDSE) for an atomic system in the radiation field, but its exact analytical solution is not known. Several numerical solutions of the Schr\"odinger equation for these kinds of systems are known \cite{cormier, chen, hansen, awasthi, chu}, but they are very time consuming for large systems and at high radiation intensities they converge slowly.   
To overcome this problem there are several theoretical approaches, which are based on the simplification of the TDSE using different approximations depending on the laser field parameters.

At low and moderate laser field intensities the time dependent perturbation theory (TDPT) is a well suited approximation for single- and multi-photon processes and even for the above threshold ionization \cite{mainfray}. In this case the TDSE is solved by considering the interaction between the laser field and the studied atomic system as a perturbation. The TDPT approach breaks down at higher laser intensities when it fails to describe the "peak suppression" in the above threshold ionization spectra \cite{kruit, xiong}.  

At higher laser field intensities other non-perturbative processes emerge like high harmonics generation, tunneling ionization (TI), over the barrier ionization (OBI), which can not be described using TDPT and other approaches are necessary.

Time independent, stationary approaches are developed based on the Floquet theory \cite{chu01}. In the framework of the generalized Floquet theory a stationary treatment of a periodic or quasiperiodic TDSE is possible. The TDSE is reduced to a set of time-independent coupled equations or into a Floquet matrix eigenvalue problem. Approaches involving the Floquet theory are applied successfully to describe multiphoton excitation, above threshold ionization and dissociation, but their main disadvantage is that in most applications the analytical form of the involved Floquet states is not known and their numerical calculation is also time consuming.

The most frequently used approaches are based on the Keldysh theory \cite{keldysh, reiss}. The Keldysh theory is based on the assumption that on the final state wavefunction only the external laser field has a dominant influence and it can be considered as a momentum eigenstate. The main shortcoming of the Keldys theory is that it completely neglects the long ranged Coulomb interaction between the ionized electron and remaining target ion. The Coulomb interaction leads to phenomena like sub-peaks in the above threshold ionization spectra \cite{freeman}, asymmetry in the spatial distribution of the ejected electron even for a symmetric few-cycle laser pulses \cite{chelk}, which can not be explained in the simplified Keldysh formalism. There are two possible ways of including the Coulomb interaction in the Keldysh formalism.  The first one is by making corrections in the transition matrix, and the second one is by making corrections in the Volkov wavefunction. Approaches based on the Keldysh theory using these corrections are applied with considerable success to study multiphoton and tunneling ionization of atomic systems \cite{reiss, popov}.

In the recent years the Coulomb-Volkov (CV) wavefunctions were used to describe processes in the presence of intense ultrashort laser field \cite{ducha01, ducha02, gayet}. The Coulomb-Volkov wavefunctions were introduced by Jain and Tzoar in 1978 \cite{tzoar} to describe laser assisted collisions. Later on they were successfully applied to describe multiphoton and above threshold ionization \cite{rodriguez, gayet}. The Coulomb-Volkov wavefunctions are also used to study ionization in tunneling and in over the barrier regime in the framework of sudden approximation \cite{ducha01, ducha02}. This model is called Coulomb-Volkov model

The accuracy of the results provided by the CV model is limited by the sudden approximation (it provides accurate results only if the pulse duration is less then two orbital period of the active electron \cite{ducha02}) and by the fact that the CV wavefunctions are only an approximate solution of the TDSE for charged particle in the presence of external radiation field.
From these two main limitations the first one is more restrictive, because it is limiting to the pulse duration in the attosecond region.
Our goal is to construct a theoretical model, which provides results as good as or better than the CV model, but which does not have the limitation in pulse duration.

Our approach is based on the approximate solution of the TDSE for quantum systems with one active electron, where the Coulomb interaction between the electron and the remaining target ion can be considered as a perturbation during the external laser pulse. The time dependent wavefunction of the active electron is expanded in terms of Volkov wavefunctions. The equation for the expansion coefficients obtained from the TDSE is solved with this approximation.

The ionization process in the over the barrier regime is considered to be a classical one and it is believed that it can be described very well by classical models like classical trajectory Monte Carlo (CTMC) \cite{olson, ducha02, hansen}.

We present double differential ionization cross sections for different laser pulses using the present, CTMC and Volkov models. The ionization probability densities are presented as a function of the electron energy and ejection angle.  

In several works \cite{mu, mishima, zhang} the effect of the Coulomb interaction on the ionization of the hydrogen atom by ultrashort laser pulses at different field intensities using different approaches is studied. In these studies the effect of the Coulomb potential is studied by employing different types of final state wavefunctions like the Coulomb-Volkov wavefunction \cite{zhang}, or first and second order Coulomb corrected Volkov wavefunctions \cite{mishima}, or by employing Coulomb corrections on the Keldysh-Faisal-Reiss theory \cite{mu}. Moreover, the effect of the Coulom potential on the overall
ionization process is also investigated.

In the present paper we  study the effect of the Coulomb potential during and after the laser pulse by analyzing the angular distribution of the electrons at given energies (extracted from the double differential cross section) and the ionization probability density (calculated from the double differential cross section by integration over the ejection angles) using various model calculations. 
The dependence of the angular distribution of photoelectrons on the ejection energy allow us to investigate the transition between a quantum type and a classical type ionization also studied by Dimitrovski and Solov'ev \cite{dimitrov} in the framework of the first Magnus approximation.  
Atomic units are used throughout the calculations. 

\section{Theory}

\subsection{Characterization of the model}
\begin{figure}
\includegraphics[width=8.5cm]{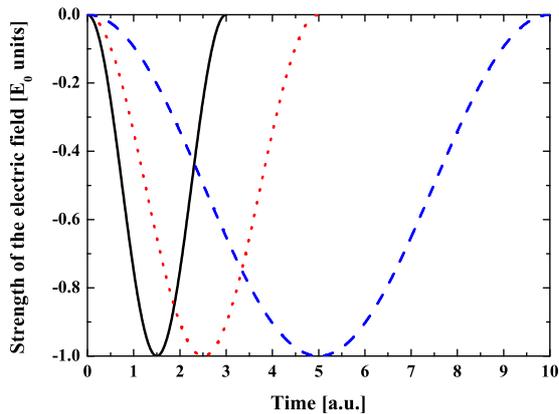}
\caption{\label{fig:1} (Color online) The electric components of the laser pulses used in our calculations. Solid line - $\tau=3$ a.u.; Dotted line - $\tau=5$ a.u.; Dashed line - $\tau=10$ a.u. }
\end{figure}

The time evolution of atomic systems in the presence of one intense ultrashort laser pulse is investigated. The laser pulse, an external electromagnetic field from the point of view of the studied system, is defined by its electric component
\begin{equation}
\vec E= \left \{
\begin{array}{ll}
 \hat\varepsilon E_0 \sin(\omega t +\phi_0)\sin^2(\frac{\pi t}{\tau})&\  \mathrm{if}\ t\in[0,\tau] \\
 0 &\ \mathrm{elsewhere}
\end{array} 
 \right.,
 \label{field}
\end{equation}
where $\hat\varepsilon$ is the polarization vector along which the laser pulse is linearly polarized, $\omega$ is the frequency of the carrier wave, $\phi_0$ is the carrier envelope phase and $\tau$ is the pulse duration. The carrier envelope phase is set as follows:
\begin{equation}
\phi_0=-\frac{\omega\tau}{2}-\frac{\pi}{2},
\end{equation} 
leading to a time symmetric laser pulse. The laser pulse is composed of a carrier wave $\sin(\omega t -\frac{\omega\tau}{2}-\frac{\pi}{2}),$ which is modulated by a sine-square envelope function $\sin^2(\frac{\pi t}{\tau})$. Figure \ref{fig:1} shows the shape of the laser pulses used in our calculations.  
%%%%%%%%%%%%%%%%%%%%%%%%figure1%%%%%%%%%%%%%%%%%%%%%%%%%%%%%%%%%%%%%%%%%%%%%%%%%%%%%%%%%%%%%%%%%%%%
%%%%%%%%%%%%%%%%%%%%%%%%figure1%%%%%%%%%%%%%%%%%%%%%%%%%%%%%%%%%%%%%%%%%%%%%%%%%%%%%%%%%%%%%%%%%%%%%

In the time evolution of the studied system one may distinguish three main time intervals.
 
In the first time interval ($t < 0$) the laser field is not yet switched on and the studied system is in a field free eigenstate. The Hamilton operator of the system in these conditions is 
\begin{equation}
\hat H_I=\frac{\hat p^2}{2}+V(\vec r),
\label{h1}
\end{equation} 
where $V(\vec r)$ is the Coulomb potential between the active electron and the rest of the system. 
The eigenvectors and eigenfunctions of this Hamilton operator are considered to be known
\begin{equation}
\hat H_I \psi_j=E_j\psi_j
\end{equation} 
and the initial state of our system coincides with eigenstate $\psi_i$. 
Under these conditions the time dependent wavefunction of the system is
\begin{equation}
\Psi_I(t)=\psi_ie^{-iE_it}.
\end{equation}

During the second time interval ($0\le t\le\tau$) the laser field is switched on and the Hamilton operator of the system becomes
\begin{equation}
\hat H_{II}=\frac{\hat p^2}{2}+\vec r\vec E+V(\vec r),
\end{equation}
where $\vec r\vec E$ is the interaction term between the laser field and the active electron expressed in $E$ gauge (Electric field gauge, length gauge). The dipole approximation was implicitly applied when the spatial dependence of $\vec E$ was neglected in Eq. (\ref{field}).
The time dependent Schr\"odinger equation 
\begin{equation}
i\frac{\partial }{\partial t}\Psi_{II}(t)=\hat H_{II}\Psi_{II}(t),
\label{tdse2}
\end{equation}
with $\Psi_{II}(t)$ time dependent wavefunction of the active electron in the presence of the laser pulse, does not have known analytical solution. In order to determine the time evolution of the studied system one needs to know the $\Psi_{II}(t)$ wavefunction. In the present approach the wavefunction is considered in the following form
\begin{equation}
\Psi_{II}(t)=\int d\vec k c(\vec k, t)\Psi_V(\vec k, t),
\label{expansion}
\end{equation}
where $\Psi_V(\vec k, t)$ are the Volkov wavefunctions. The Volkov wavefunctions are the solutions of the TDSE in dipole approximation for a charged particle in radiation field 
\begin{equation}
i\frac{\partial}{\partial t}\Psi_V(\vec k, t)=\left(\frac{\hat p^2}{2}+\vec r\vec E\right)\Psi_V(\vec k,t),
\label{tdsevolkov}
\end{equation}
and they can be expressed as
\begin{equation}
\Psi_V(\vec k, t)=e^{-\frac{i}{2}\int_0^tdt'(\vec k+\vec A(t'))^2}e^{i(\vec k+\vec A(t))\vec r}.
\label{volkov}
\end{equation} 
In the above expression 
\begin{equation}
\vec A(t)=-\int_0^t\vec E (t')dt'
\end{equation}
is the vector potential of the electromagnetic field.

The time dependent wavefunction is well defined by the expansion coefficients $c(\vec k, t)$, so if one is interested in the time evolution of the system is enough to calculate these coefficients.

By substituting the time dependent wavefunction of Eq. (\ref{expansion}) into the TDSE given by Eq. (\ref{tdse2}) can be obtained 
\begin{equation}
i\frac{\partial}{\partial t}\int d\vec k c(\vec k,t) \Psi_V(\vec k,t)=\left(\frac{\hat p^2}{2}+V(\vec r)+\vec r\vec E \right)\int d\vec k c(\vec k,t)\Psi_V(\vec k, t).
\label{tdse21}
\end{equation}
Using Eq. (\ref{tdsevolkov}) it is easy to prove that
\begin{equation}
\int d\vec k c(\vec k,t)i\frac{\partial}{\partial t}\Psi_V(\vec k, t)=\left(\frac{\hat p^2}{2}+\vec r\vec E \right)\int d\vec k c(\vec k,t)\Psi_V(\vec k, t),
\end{equation}
which can be used to simplify Eq. (\ref{tdse21})
\begin{equation}
i\int d\vec k \Psi_V(\vec k,t)\frac{\partial}{\partial t}c(\vec k,t)=\int d\vec k c(\vec k,t) V(\vec r)\Psi_V(\vec k, t).
\label{koz}
\end{equation}
Eq. (\ref{koz}) can be converted into a more favorable form by transforming it into momentum space
\[
i\int d\vec k \left(\frac{\partial}{\partial t}c(\vec k,t)\right)e^{-\frac{i}{2}\int_0^tdt'(\vec k+\vec A(t'))^2}\int d\vec r e^{i(\vec k-\vec p+\vec A(t))\vec r}=
\]
\begin{equation}
\int d\vec k c(\vec k,t)e^{-\frac{i}{2}\int_0^tdt'(\vec k+\vec A(t'))^2}\int d\vec r V(\vec r) e^{i(\vec k-\vec p+\vec A(t))\vec r}.
\end{equation}
After basic mathematical operations and the substitution $\vec q = \vec p -\vec A(t)$, this equation becomes
\begin{equation}
i(2\pi)^3\frac{\partial}{\partial t}c(\vec q,t)e^{-\frac{i}{2}\int_0^tdt'(\vec q+\vec A(t'))^2}=\int d\vec k c(\vec k,t)e^{-\frac{i}{2}\int_0^tdt'(\vec k+\vec A(t'))^2}\int d\vec r V(\vec r) e^{i(\vec k-\vec p+\vec A(t))\vec r}.
\label{diff1}
\end{equation}
By introducing the notation $\vec s=\vec k-\vec q$ and by restructuring Eq. (\ref{diff1}) for the expansion coefficients $c(\vec k,t)$ can be obtained the following integro-differential equation
\begin{equation}
\frac{\partial}{\partial t} c(\vec q,t)=-\frac{i}{(2\pi)^3}\int d\vec s c(\vec s+\vec q,t)e^{-\frac{i}{2}\int_0^tdt'\vec s\left(\vec s+2\vec q+2\vec A(t')\right)}\int\vec drV(\vec r)e^{i\vec s\vec r}.
\label{diff2}
\end{equation}

We note, however, that Eq. (\ref{diff2}) is equivalent with the Schr\"odinger equation given by Eq. (\ref{tdse2}). By solving Eq. (\ref{diff2}) one obtains directly the time dependent wavefunction in momentum space, which carries all the information about the studied system. Eq. (\ref{diff2}) can be solved numerically, but this direct approach needs large computational resources, and in several cases it is not worth to do it, because by introducing some approximations one may keep almost the same accuracy.

In the third time interval ($t > \tau$) the laser field is switched off and the Hamilton operator of the system can be expressed as
\begin{equation}
\hat H_{III}=\frac{\hat p^2}{2}+V(\vec r),
\end{equation}
which is identical with the one given in the first time interval.\\
The time dependent wavefunction of the system in this time interval can be given as a linear combination of stationary-state wavefunctions
\begin{equation}
\Psi_{III}(\vec r,t)=\sum_b\psi_be^{-iE_bt}+\int d\vec k c_f(\vec k)\psi_fe^{-iE_ft}, 
\end{equation}
where $\psi_b$ represents bound, while $\psi_f$ represents free states. For convenience the free states are represented by plane waves, so we have
\begin{equation}
\psi_f=e^{i\vec k\vec r}
\end{equation}
and
\begin{equation}
E_f=\frac{k^2}{2}.
\end{equation}

One of the basic properties of the time dependent wavefunction, which describes the evolution of a real system is that it is continuous over time.

From the continuity condition at time $t=0$ one obtains 
\begin{equation}
\Psi_{I}(t=0)=\Psi_{II}(t=0),
\end{equation}
which has the following explicit form 
\begin{equation}
\psi_i=\int d\vec k c(\vec k,t=0)e^{i\vec k\vec r}.
\end{equation}
From this expression one may get the initial condition for Eq. (\ref{diff2})
\begin{equation}
c(\vec q,t=0)=\frac{1}{(2\pi)^3}\langle e^{i\vec q\vec r}|\psi_i\rangle
\label{initcond}
\end{equation}
It is worth to notice that the above initial condition is in fact the initial state wavefunction in the momentum space, because at $t=0$ the Volkov wavefunctions are reduced to plane waves. 

From continuity condition at $t=\tau$ we can write:
\begin{equation}
\Psi_{II}(t=\tau)=\Psi_{III}(t=\tau),
\end{equation}
which can be given explicitly as
\begin{equation}
\int d\vec k c(\vec k,\tau)e^{-\frac{i}{2}\int_0^\tau dt'(\vec k+\vec A(t'))^2}e^{i\vec r(\vec k+\vec A(\tau))}=\sum_b\psi_be^{-iE_b\tau}+\int d\vec kc_f(\vec k)e^{i(\vec k\vec r-E_f\tau)}.
\end{equation}

We are interested in the transition probability to a free final state, which means that we need to know $c_f(\vec k)$. This can be obtained by transforming the above equation into momentum space and by neglecting the contribution of bond states
\begin{equation}
c_f(\vec p)=c(\vec p-\vec A(\tau),\tau)e^{-i[E(\vec p-\vec A(\tau),\tau)-E_f\tau]},
\label{final}
\end{equation}
where we used the notation
\begin{equation}
E(\vec k,T)=\frac{1}{2}\int_0^Tdt'(\vec k+\vec A(t'))^2.
\end{equation}
The $c_f(\vec p)$ given by Eq. (\ref{final}) is the final-state wavefunction in momentum space.\\
The transition probability from the initial state $\psi_i$ to a free final state $\psi_f$, with a well defined momentum $\vec p$, is given as
\begin{equation}
P_{i\rightarrow f}= (2\pi)^3\left|c_f(\vec p) \right|^2=(2\pi)^3\left|c(\vec p-\vec A(\tau),\tau) \right|^2.
\label{ionprob}
\end{equation}

%%%%%%%%%%%%%%%%%%%%%%%%%%%%%%%%%%%%%%%%%%%%%%%%%%%%%%%%%%%%%%%%%%%
\subsection{Volkov model as a zero order approximation }
%%%%%%%%%%%%%%%%%%%%%%%%%%%%%%%%%%%%%%%%%%%%%%%%%%%%%%%%%%%%%%%%%%%

The critical point of our model is how to solve Eq. (\ref{diff2}) or in other words how to propagate our system in the presence of external laser field in the second time interval. The simplest way is by neglecting completely the Coulomb potential ($V(\vec r)=0$). This approximation provides good results only for high laser field intensities. In this framework Eq. (\ref{diff2}) becomes
\begin{equation}
\frac{\partial}{\partial t} c(\vec q,t)=0,
\end{equation}    
which has the following analytical solution using the initial condition (\ref{initcond})
\begin{equation}
c(\vec q,t)\equiv c^{(0)}(\vec q)=\frac{1}{(2\pi)^3}\langle e^{i\vec q\vec r}|\psi_i\rangle.
\end{equation}
From Eq. (\ref{final}) the final state wavefunction in momentum space can be written as:
\begin{equation}
c_f(\vec p)=c^0(\vec p-\vec A(\tau))e^{-i[E(\vec p-\vec A(\tau),\tau)-E_f\tau]},
\label{volkovmodel}
\end{equation}
which is the initial state wavefunction in the momentum space shifted by the momentum transfer $\vec A(\tau)$ gained from the external laser field.

%%%%%%%%%%%%%%%%%%%%%%%%%%%%%%%%%%%%%%%%%%%%%%%%%%%%%%%%%%%%%%%%%%%%%
\subsection{First order approximation }
%%%%%%%%%%%%%%%%%%%%%%%%%%%%%%%%%%%%%%%%%%%%%%%%%%%%%%%%%%%%%%%%%%%%%%%
In most cases the Volkov model (see Eq. (\ref{volkovmodel})) does not provide accurate results, because the Coulomb interaction at moderate intensities can not be totally neglected. At moderate laser intensities one can assume that the influence of the Coulomb interaction on the evolution of the system in the second time interval is small and the expansion coefficients $c(\vec k, t)$ are close to that ones provided by the Volkov model (\ref{volkovmodel}). Based on this arguments Eq. (\ref{diff2}) can be simplified by replacing $c(\vec k, t)$ in the right hand side by $c^{(0)}(\vec k)$ as follows
\begin{equation}
\frac{\partial}{\partial t} c^{(1)}(\vec q,t)=-\frac{i}{(2\pi)^3}\int d\vec s c^{(0)}(\vec s+\vec q)e^{-\frac{i}{2}\int_0^tdt'\vec s\left(\vec s+2\vec q+2\vec A(t')\right)}\int\vec drV(\vec r)e^{i\vec s\vec r}.
\label{firstorder}
\end{equation}
The same equation can be obtained using a different approach by considering the Coulomb interaction as a small perturbation and by keeping only the first order terms in $V(\vec r)$. 
The advantage of this approximation is that it eliminates the direct coupling between the expansion  coefficients $c(\vec k, t)$, which makes easier and faster the solution of Eq. (\ref{diff2}).\\
Eq. (\ref{firstorder}) can be simplified and its solution can be given as simple integral
\begin{equation}
c^{(1)}(\vec q,t)=c^{(0)}(\vec q)-\frac{i}{(2\pi)^3}\int_0^t dt' I(\vec q, t'), 
\label{diff3}
\end{equation}  
where
\begin{equation}
I(\vec q, t)=\int d\vec s c^{(0)}(\vec s+\vec q)e^{-\frac{i}{2}\int_0^tdt'\vec s\left(\vec s+2\vec q+2\vec A(t')\right)}\int\vec drV(\vec r)e^{i\vec s\vec r}.
\label{inteq}
\end{equation}

%%%%%%%%%%%%%%%%%%%%%%%%%%%%%%%%%%%%%%%%%%%%%%%%%%%%%%%%%%%%
\subsection{The Classical Trajectory Monte Carlo simulation}
%%%%%%%%%%%%%%%%%%%%%%%%%%%%%%%%%%%%%%%%%%%%%%%%%%%%%%%%%%%%
 
In the last two decades there was a great revival of the classical trajectory Monte Carlo (CTMC) calculations applied in atomic collisions involving three or more particles  \cite{Olson91}. This approximation seems to be useful in treating atomic collisions where the quantum mechanical ones become very complicated or unfeasible. This is the case usually when higher order perturbations should be applied or many particles take part in the processes. The CTMC method has been quite successful in dealing with the ionization process in laser-atom collisions, when, instead of the charged particles, electromagnetic fields are used for excitation 
of the target. The CTMC method is a nonperturbativ method, where classical equations of motions are solved numerically. 
A micro-canonical ensemble characterizes the initial state of the  
target. In this work, the initial conditions of the target are taken from this ensemble, which is constrained to an initial binding energy of H(1s) (0.5 a.u.).  

In the present CTMC approach, Newton's classical nonrelativistic equations of motions are solved \cite{Abrines66,Olson77,Tokesi94} numerically when an external laser field given by Eq. (\ref{field}) is included.
For the given initial parameters, Newton's equations of motion were integrated with 
respect to time as an independent variable by 
the standard Runge-Kutta method until the real exit channels are obtained.   
For the ionization channel the final energy and the scattering angles 
(polar and azimuth) of the projectile and the ionized electron were recorded. 
These parameters were calculated at large separation of the ionized electron
and the target nucleus, where the Coulomb interaction is negligible.

The single and double differential ionization probabilities ($P_i$) were computed 
with the following formulas:

\begin{equation}
{{dP_i}\over{dE}}={{N_i}\over{N \Delta E 
}},
\label{eq:engdiff}
\end{equation}

\begin{equation}
{{dP_i}\over{d\Omega}}={{N_i}\over{N 
\Delta \Omega}},
\label{eq:angdiff}
\end{equation}

\begin{equation}
{{d^2P_i}\over{dEd\Omega}}={{N_i}\over{N 
\Delta E \Delta \Omega}}.
\label{eq:doublediff}
\end{equation}

\noindent The standard deviation for a differential probabilities is defined through:
 
\begin{equation}
\Delta P_i= P_i \left [{{N-N_i}\over{N-N_i}}
\right ] ^{1/2}.
\label{eq:err}
\end{equation}
 
In Eqs. (\ref{eq:engdiff}- \ref{eq:err}) $N$ is the total number of trajectories calculated for the given collision system, $N_{i}$  is the number of trajectories that satisfy the criteria for ionization under consideration in the energy interval $\Delta E$ and the emission angle interval $\Delta \Omega$ of the electron.

%%%%%%%%%%%%%%%%%%%%%%%%%%%%%%%%%%%%%%%%%%%%%%%%%%%%%%%%%%%%%%%%%%%%%%%%%%%
\section{Ionization of the hydrogen atom}
%%%%%%%%%%%%%%%%%%%%%%%%%%%%%%%%%%%%%%%%%%%%%%%%%%%%%%%%%%%%%%%%%%%%%%%%%%%

The theoretical approaches presented above are applied to describe the ionization of the hydrogen atom in over the barrier regime. We choose this system because the calculations are easy to perform and because there are several theoretical studies on this system \cite{hansen, ducha01, ducha02, mishima, zhang} and it is easy to compare our results with others found in the literature \cite{ducha02}.\\
Using the 1s orbital of the hydrogen atom as initial state wavefunction
\begin{equation}
\psi_i=\frac{1}{\sqrt{\pi}}e^{-r}
\end{equation}
one obtains the following initial condition for Eq. (\ref{diff2})
\begin{equation}
c^{(0)}(\vec q) = \frac{1}{\pi^2\sqrt \pi (1+\vec q^2)^2}.
\label{inithyd}
\end{equation}
Using Eq. (\ref{inithyd}) the ionization probability in Volkov approximation can be expressed as
\begin{equation}
P_{i\rightarrow f}(\vec p)=\frac{16}{\pi(1+(\vec p-\vec A(\tau))^2)^4}.
\end{equation}
The first order model can also be adapted very easily by using the Coulomb potential
\begin{equation}
V(\vec r)=-\frac{1}{r}
\end{equation}
in Eq.(\ref{inteq}), which can be significantly simplified by performing some of the involved integrals
\[
I(\vec q, t)=\lim_{r_{max}\rightarrow +\infty}\frac{8}{\sqrt \pi}\int_0^\infty\int_0^\pi\frac{\sin(\theta_s)d\theta_sds}{(1+s^2+q^2+2sq\cos(\theta_s))^2}\quad\times
\]
\begin{equation}
e^{-\frac{i}{2}(s^2t+2sq\cos(\theta_s)t+2f(t)s\cos(\theta_s)\cos(\theta_\varepsilon))}J_0(2sf(t)\sin(\theta_s)\sin(\theta_\varepsilon))(\cos(sr_{max})-1),
\label{firstsol}
\end{equation}
where $J_0$ is the Bessel function of the first kind.
The angle between $\vec s$ and $\vec q$ is $\theta_s$, while the angle between $\hat \varepsilon$ and $\vec q$ is $\theta_\varepsilon$ and  
\begin{equation}
f(t)=\int_0^{t}A(t')dt'.
\end{equation}
By substituting  Eq.(\ref{firstsol}) into Eq.(\ref{diff3}) and by performing the remaining integrals numerically one obtains the expansion coefficient in first order approximation. From the obtained expansion coefficients the ionization probability is calculated by using Eq.(\ref{ionprob}).

%%%%%%%%%%%%%%%%%%%%%%%%%%%%%%%%%%%%%%%%%%%%%%%%%%%%%%%%%%%%%%%%%%%%%%%%%%%%%%%%%%%%%
\section{Results and discussion}
%%%%%%%%%%%%%%%%%%%%%%%%%%%%%%%%%%%%%%%%%%%%%%%%%%%%%%%%%%%%%%%%%%%%%%%%%%%%%%%%%%%%%

Calculations are performed using laser pulses with duration $\tau$ of $3$ a.u., $5$ a.u. and $10$ a.u. at two different field intensities ($E_0=1$ a.u. and $E_0=10$ a.u) for each pulse duration. The shape of the pulses can be seen on FIG. \ref{fig:1}.\\
%%%%%%%%%%%%%%%%%%%%%%%%%%%%%%%figure2%%%%%%%%%%%%%%%%%%%%%%%%%%%%%%%%%%%%%%%%%%%%%      
\begin{figure}
%two column figure
\includegraphics[width=16cm]{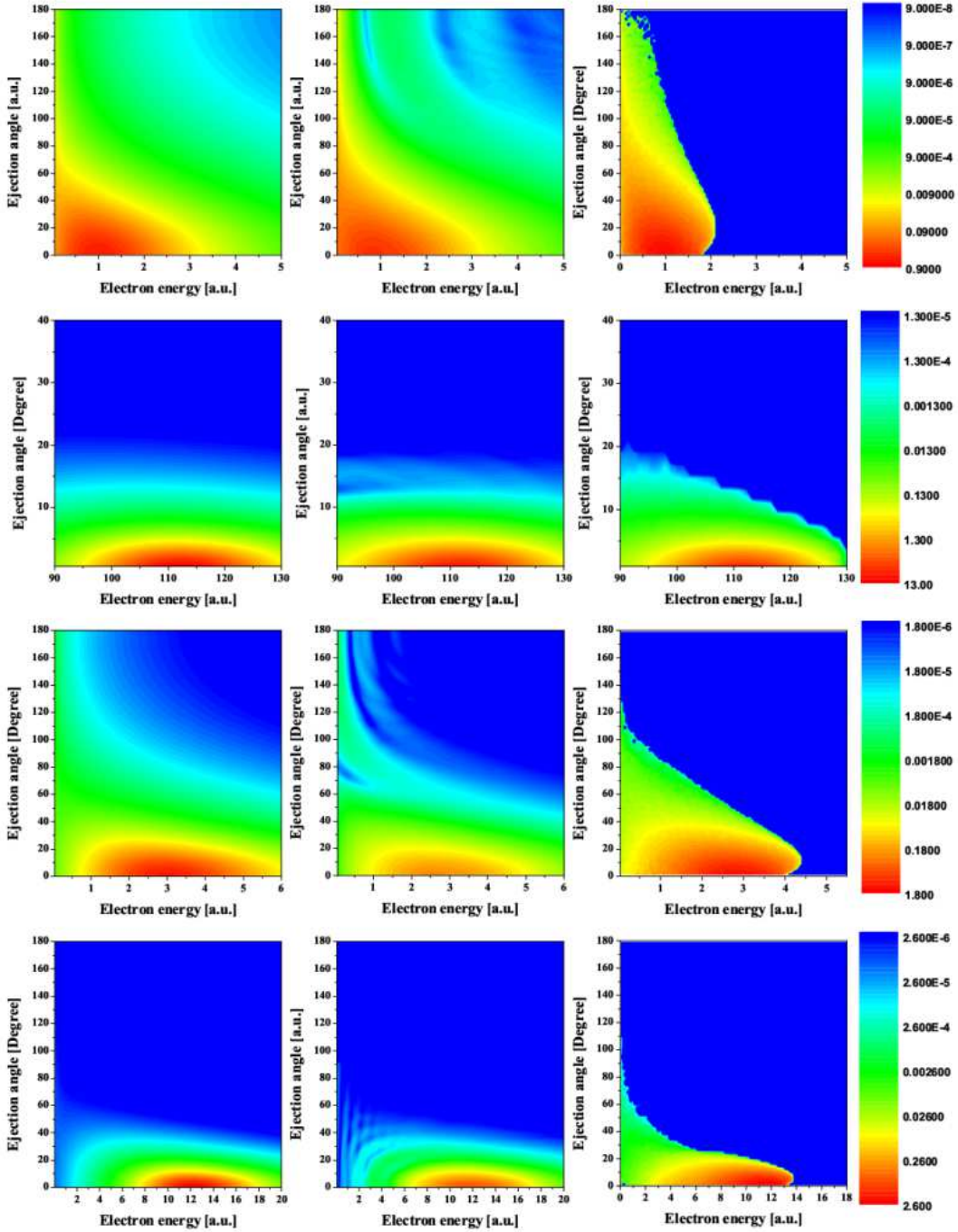}
\caption{\label{fig:2} (Color online) Two-dimensional ionization probability density in the yOz positive semi-plane as a function of the  electron energy and the ejection angle. First column - Volkov results; Second column - first order results; Third column - CTMC results; First row - $\tau = 3$ a.u. and $E_0 = 1$ a.u.; Second row - $\tau = 3$ a.u. and $E_0 = 10$ a.u.; Third row - $\tau = 5$ a.u. and $E_0 = 1$ a.u.; Fourth row - $\tau = 10$ a.u. and $E_0 = 1$ a.u.}
\end{figure}
%%%%%%%%%%%%%%%%%%%%%%%%%%%%%%%figure2%%%%%%%%%%%%%%%%%%%%%%%%%%%%%%%%%%%%%%%%%%%%%%
The double differential ionization cross sections calculated using the Volkov, the first order and the CTMC model are presented on Figure \ref{fig:2}, where the ionization probability density is plotted as a function of the electron energy and the ejection angle. The ejection angle is measured from the polarization vector $\hat\varepsilon$, which is in the direction of z axis. Due to the spherical symmetry of the ground state wavefunction and of the Coulomb potential the double differential ionization cross section has a rotational symmetry, where the rotational axis is in the direction of the polarization vector $\hat\varepsilon$ of the laser field. Due to this symmetry the yOz positive semi-plane carries all the information about the ionization cross section. At first sight one can observe that at a large scale all three models predicts the same cross sections. In each approach the electrons are ejected with maximum probability along the polarization vector $\hat\varepsilon$ with energy around the value $\frac{A(\tau)^2}{2},$ which is gained by the momentum transfer $\vec A(\tau)$ from the external laser field. 

After detailed analysis, important differences can be observed. In the case of the first order and CTMC model the maxima of the predicted cross sections are shifted toward smaller energies. This shift is caused by the Coulomb attraction during the ionization between the active electron and the rest of the system, in a classical picture the ejected electrons are decelerated by the Coulomb attraction.    
%%%%%%%%%%%%%%%%%%%%%%%%%%%%%%%%%%%%figure3%%%%%%%%%%%%%%%%%%%%%%%%%%%%%%%
\begin{figure}
%two column figure
\includegraphics[width=14cm]{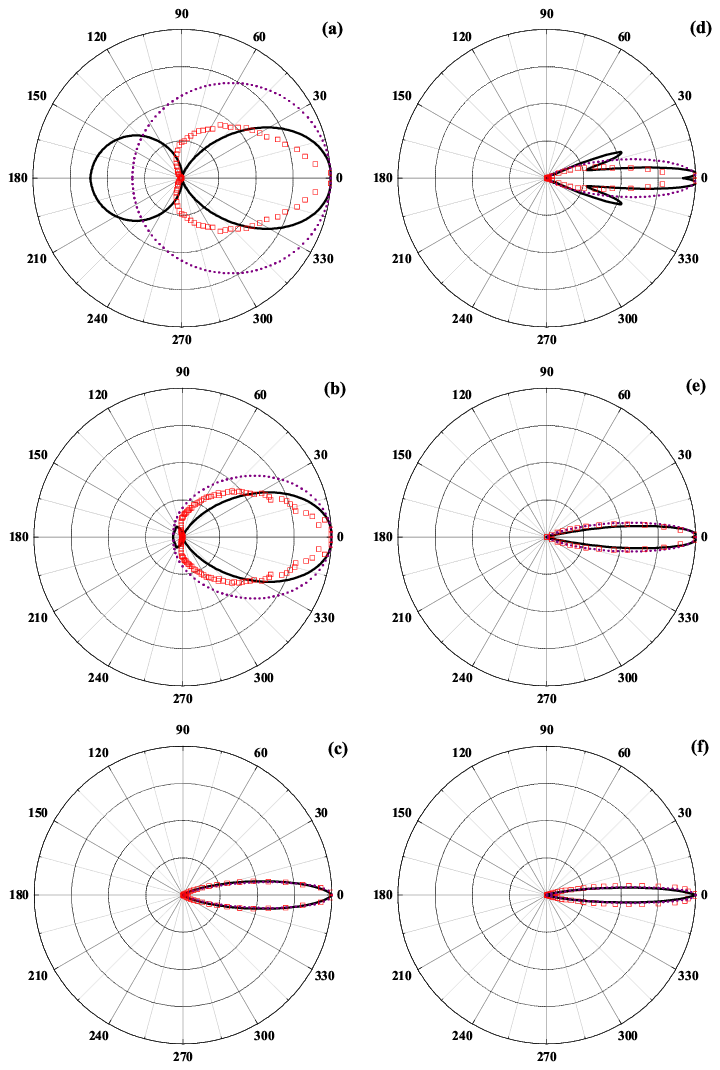}
\caption{\label{fig:3} (Color online) Angular distribution of the photoelectrons for given electron energies ($W_e$). Solid line - first order; Dotted line - Volkov; Squares - CTMC; First column - $\tau = 5$ a.u. and $E_0 = 1$ a.u.; Second column - $\tau = 10$ a.u. and $E_0 = 1$ a.u.; (a) $W_e = 0.02$ a.u.; (b) $W_e = 0.125$ a.u.; (c) $W_e = 3.5 $ a.u.; (d)  $W_e = 4.5$ a.u.; (e) $W_e = 6$ a.u.; (f)  $W_e = 10$ a.u.}
\end{figure}
%%%%%%%%%%%%%%%%%%%%%%%%%%%%%%%%%%%%figure3%%%%%%%%%%%%%%%%%%%%%%%%%%%%%%%%%%%%%%%%%%%%%%%%%%%%%%%

Other important differences can be identified at the angular distribution of the ejected electrons at fixed energies ($W_e$) as shown in Figure \ref{fig:3}. The maximum value of each angular distribution is normalized to unity to allow an easier comparison. 

At the low part of the energy spectrum (see FIG. \ref{fig:3} (a) and (d), with $W_e$ below the peak observable on FIG. 2) significant differences are observed between the predicted first order and CTMC angular distributions. The observed differences implies the quantum nature of the ionization for low energy electrons, which can not be described correctly by classical calculations. These differences start to disappear at higher electron energies (see FIG. \ref{fig:3} (b) and (e)), where both distributions roughly are the same. However some minor differences still exist showing a transition between the quantum and classical nature of the ionization. At high energies, the differences observed at lower energies completely disappear (see FIG. \ref{fig:3} (e) and (f)) and the distributions predicted by first order and CTMC model are in good agreement indicating that the ionization mechanism for electrons ejected at high energies behave classically. This evolution of the cross section and the transition between the quantum and classical type ionization was also studied by Dimitrovski and Solov'ev \cite{dimitrov} in the framework of first Magnus approximation. They obtained the same transition between quantum and classical type ionization by varying the net momentum transfer.

At the low part of the energy spectrum (see FIG. \ref{fig:3} (a) and (d)) large discrepancies are also observed between the predicted first order and Volkov angular distributions. These discrepancies show the significant influence of the Coulomb interaction on the ionization of the low energy electrons. This disagreement between the discussed models diminishes and completely disappears at higher electron energies (see FIG. \ref{fig:3} (b, c, e, f)), which indicates that the Coulomb interaction has less influence on the ionization of the high energy electrons.   
%%%%%%%%%%%%%%%%%%%%%%%%%%%%%%%%%figure4%%%%%%%%%%%%%%%%%%%%%%%%%%%
\begin{figure}
%one column figure
\includegraphics[width=8.5cm]{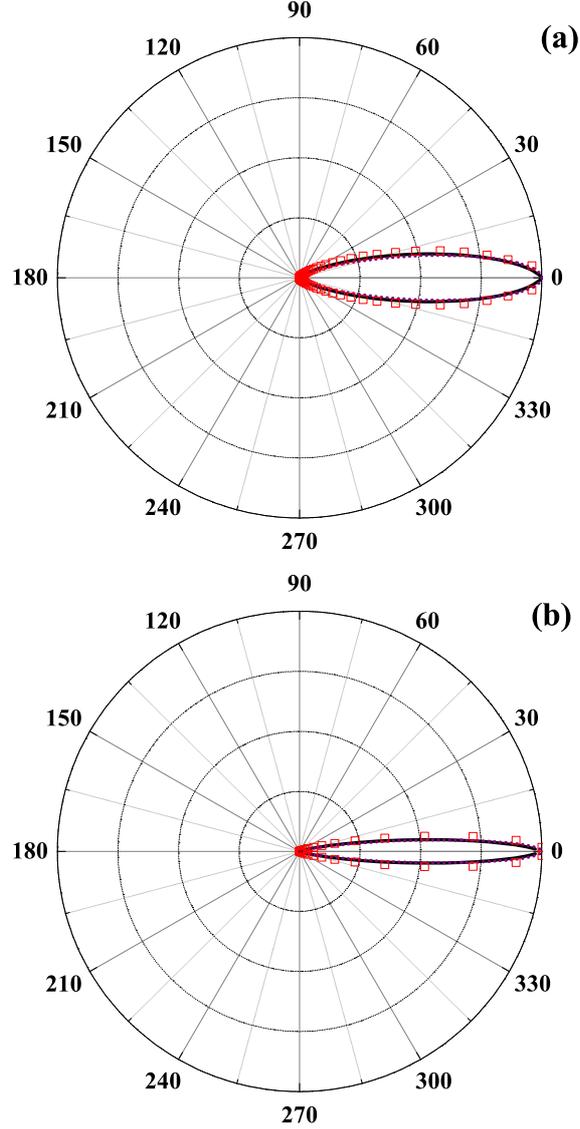}
\caption{\label{fig:4} (Color online) The net angular distribution of the photoelectrons. Solid line - first order; Dotted line - Volkov; Squares - CTMC; (a) - $\tau = 5$ a.u. and $E_0 = 1$ a.u.; (b) - $\tau = 10$ a.u. and $E_0 = 1$ a.u.}
\end{figure}
%%%%%%%%%%%%%%%%%%%%%%%%%%%%%%%%%figure4%%%%%%%%%%%%%%%%%%%%%%%%%%%
Figure \ref{fig:4} shows the net angular distributions of the ejected electrons obtained from the double differential cross section by integrating over the electron energies for different pulse durations. The angular distributions predicted by different models are  in good agreement with each other, because the contribution of the low energy electrons in the net angular distribution is negligible, so the influence of the Coulomb potential and the quantum nature of the ionization can be  also neglected.  
%%%%%%%%%%%%%%%%%%%%%%%%%%%%%figure5%%%%%%%%%%%%%%%%%%%%%%%%%%%%%%%%%
\begin{figure}
% two column figure
\includegraphics[width=15cm]{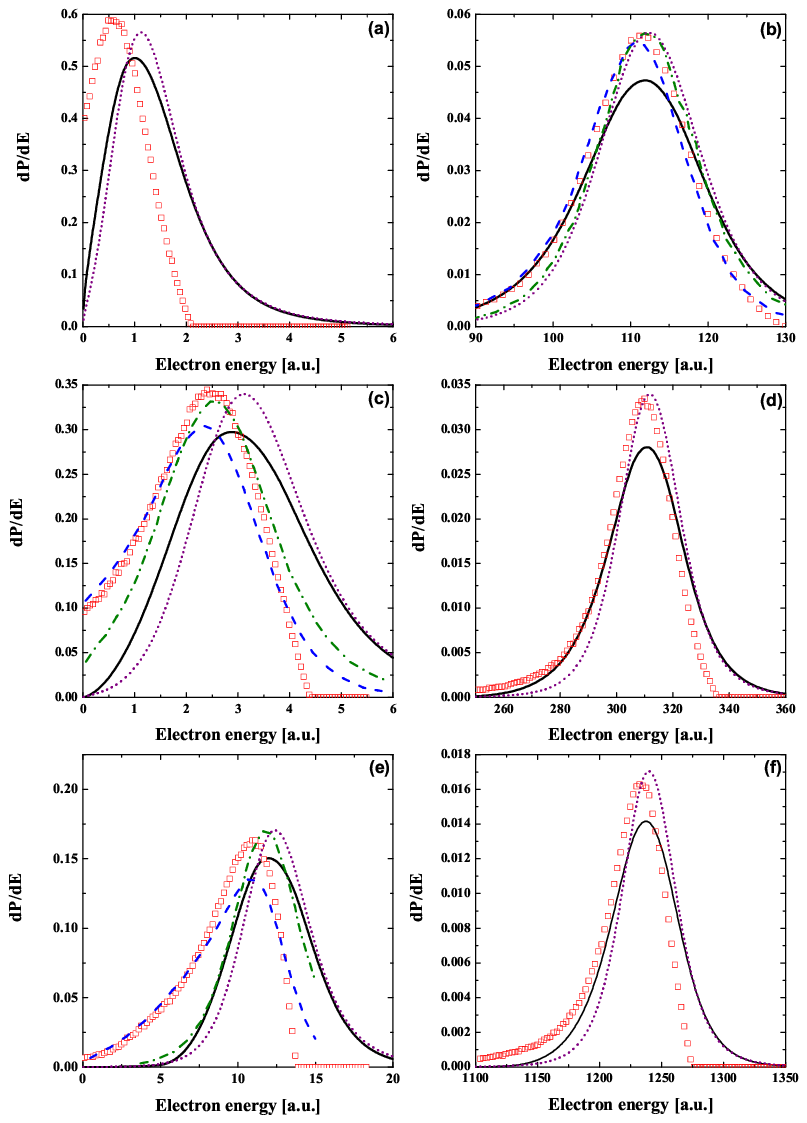}
\caption{\label{fig:5} (Color online) Ionization probability density as function of the electron energy. Solid line - first order; Dotted line - Volkov; Squares - CTMC; Dashed line - TDSE \cite{ducha02}; Dash-dotted line - CV \cite{ducha02}  First column - $E_0 = 1$ a.u.; Second column - $E_0 = 10$ a.u.; First row - $\tau = 3$ a.u.; Second row - $\tau = 5$ a.u.; Third row - $\tau = 10$ a.u.}
\end{figure}
%%%%%%%%%%%%%%%%%%%%%%%%%%%%%figure5%%%%%%%%%%%%%%%%%%%%%%%%%%%%%%%%%
On Figure \ref{fig:5} we have represented the ionization probability densities obtained from the double differential ionization cross sections by integrating over the ejection angle. The  probability densities are calculated using the first order, the Volkov and CTMC models. Figure \ref{fig:5} also shows the TDSE and CV results calculated by Duchateau et. al. \cite{ducha02}.
At high laser field intensity ($E_0=10$ a.u.) a very good agreement is observed between TDSE and CTMC results on Figure \ref{fig:5} (b). At lower laser intensity ($E_0=1$ a.u.) the agreement is acceptable (see FIG. \ref{fig:5} (b) and (c) ), but not as good as in the previous case, because at these intensities tunneling ionization also takes place, which is not considered in our CTMC model. This good agreement between classical and quantum approaches at high laser field intensities was also confirmed by several other studies \cite{hansen, ducha02}, which proves the classical nature of the over the barrier ionization.

\begin{figure}
% one column figure
\includegraphics[width=8.5cm]{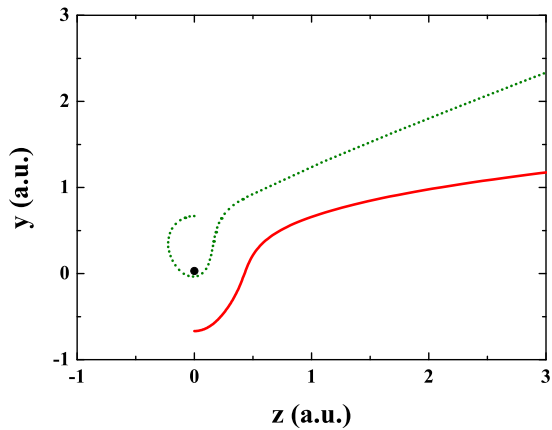}
\caption{\label{fig:6}(Color online) Two possible electron trajectories during the ionization corresponding to photoelectrons with high (Solid line) and low (Dotted line) ejection energies. The black dot is the residual proton and the net momentum transfer is along the z axis.}
\end{figure}

The main difference between the Volkov and the first order model is that the Volkov model completely neglects the Coulomb attraction during the ionization process. The effect of the Coulomb interaction on the photoelectrons during the ionization can be studied by comparing the Volkov and first order ionization probability densities. At the high energy part of each photoelectron spectrum presented on Figure \ref{fig:5}, a good agreement can be observed between the Volkov and first order model. This agreement breaks down at the lower energy part of the spectra, where the first order spectrum is shifted toward the lower energies. The above behavior was also observed in the angular distribution of ionized electrons and it can be explained using a very simple intuitive picture presented on Figure \ref{fig:6}. The electrons with high ejection energy have their momentum in the initial state in the same direction as the net momentum transfer. In this way the trajectory of the high energy electrons leads directly away from the core, with a very small portion close to the core where the Coulomb potential has a significant influence. On the other hand the electrons with lower ejection energies have a momentum in the initial state, which leads in the opposite direction of the net impulse transfer. In this way the low energy electrons need to "go around" the core, leading to a trajectory, which has a long portion close to the core, where they can be influenced significantly by the Coulomb interaction.

%%%%%%%%%%%%%%%%%%%%figure6%%%%%%%%%%%%%%%%%%%%%%%%%%%%%%%%%%%%%%%%%%
%%%%%%%%%%%%%%%%%%%%figure6%%%%%%%%%%%%%%%%%%%%%%%%%%%%%%%%%%%%%%%%%%

Results obtained by Duchateau et. al. \cite{ducha02} using Coulomb-Volkov wavefunctions and the sudden
approximation is also included in our analysis. In the CV model the influence of the Coulomb interaction
on the ionization process is not included, it is taken into account only in the final state wavefunction.
The comparison of CTMC, CV and first order results allow us to study in detail the influence of the Coulomb
interaction during and after the laser pulse.

At low intensities for short pulse, the agreement between the CV and CTMC results is better than between the CTMC and first order results (see FIG. \ref{fig:5} (a)). For this pulse parameters the net momentum transfer is small leading to photoelectrons with small momentum, which can not be described correctly by the plane waves employed by the first order model. 

At the same intensity but for longer pulses, major discrepancies can be observed between CV and CTMC results (see FIG. \ref{fig:5} (c, e)), because for these pulse durations the sudden approximation no longer applies. The same discrepancies are observed between first order and CTMC results, because the final state wavefunction is still inaccurate.  

At high laser field intensities the agreement between the first order and CTMC results is better than the agreement between CV and CTMC results (see FIG. \ref{fig:5} (b)). In this case the momentum transfer is high leading to photoelectrons with high energies, where the plane wave approximation for the final state is accurate. At the low energy part of the spectra presented on Figure \ref{fig:5} (b, d, f) the first order results show good agreement with the CTMC results (considered identical with the TDSE results at this high intensity), but discrepancies exists at the high energy part of the spectra. Both first order and CV models predict a higher ionization probability than the CTMC model at the high energy part of the spectra, which in the case of the first order model can be corrected by taking into account the Coulomb interaction after the laser pulse.  

%%%%%%%%%%%%%%%%%%%%%%%%%%%%%%%%%%%%%%%%%%%%%%%%%%%%%%%%%%%%%
\section{Conclusions and Outlook}
%%%%%%%%%%%%%%%%%%%%%%%%%%%%%%%%%%%%%%%%%%%%%%%%%%%%%%%%%%%%%

In the present work a general approach for the ionization of atomic systems by ultrashort laser pulses in one active electron approximation was presented. The time dependent Schr\"odinger equation of the studied system was transformed into a pseudo-momentum space, where it was solved in an iterative way. In the zero order solution the Coulomb potential was neglected (Volkov model), while in the first order solution it was taken into account during the ionization as a perturbation.

Calculations were performed for the ionization of the hydrogen atom in over the barrier regime using the Volkov and the first order solution of the TDSE. Classical trajectory Monte-Carlo calculations were also performed. The obtained results were analyzed and compared with results obtained by Duchateau and coworkers \cite{ducha02} using TDSE and CV model.

The double differential ionization cross sections were calculated using Volkov, first order and CTMC models for different laser pulses. A good agreement was found between the results using these three models at high laser field intensities. This good agreement between classical and  quantum calculations was also confirmed by other groups \cite{hansen, ducha02}. At lower intensities, however, small discrepancies appeared due to the tunneling ionization.

The first order and CTMC results were shifted toward smaller energies due to the Coulomb attraction of the remaining target ion. 

More differences were identified by analyzing the angular distribution of the ejected electrons.
For low energy electrons major discrepancies have been found between CTMC and first order results, because the ionization of this low energy electrons is a quantum process, which can not be described in a classical approach. 
At higher electron energies, the observed discrepancies disappear leading to the conclusion that the ionization process of the high energy electrons can be described classically. 
At low energies, differences were also observed between the first order and Volkov results, which disappeared at higher electron energies indicating that during the ionization electrons with higher energies are less influenced by the Coulomb interaction. The same observation was made by analyzing the ionization probability densities, where significant differences between the Volkov and first order results were observed for low energy photoelectrons. A simple explanation for this behavior was found by considering the possible classical trajectories of the electrons during the ionization. Similar conclusion was also drawn by Zhang et. al. \cite{zhang} in their study, where they found that the influence of the Coulomb interaction depends mainly on the energy of the photoelectrons rather than on the laser field intensities. 

The influence of the Coulomb potential during and after the laser pulse was studied. For short laser pulses with low momentum transfer we found that the Coulomb potential has an important role after the laser pulse is switched off. In the case of longer pulses, with low momentum transfer, the Coulomb potential is equally important during and after the pulse, while in the case of high momentum transfer the Coulomb interaction has a greater influence during the laser pulse. 

A good agreement between first order and CTMC results was found at high laser field intensities, where the momentum was high, while at lower intensities with low momentum the agreement was acceptable, comparable with the agreement between the CV and CTMC results.

Our model can be further improved by considering the influence of the Coulomb potential on the final state wavefunction, by using Coulomb wavefunctions instead of simple plane waves for the final state, which will be the subject of our future investigations.

%%%%%%%%%%%%%%%%%%%%%%%%%%%%%%%%%%%%%%%%%%%%%%%%%%%%%%%%%%%%%%%
\section{Acknowledgment}
%%%%%%%%%%%%%%%%%%%%%%%%%%%%%%%%%%%%%%%%%%%%%%%%%%%%%%%%%%%%%%

The work was supported by the Romanian Academy of Sciences (grant No. 35/3.09.2007), the Romanian National Plan for Research (PN II) under contract No. ID$\_539$, the grant ``Bolyai'' from the Hungarian Academy of Sciences, the Hungarian National Office for Research and Technology.

\end{document}